\documentclass[aps,prl,twocolumn]{revtex4-1}
\usepackage{graphicx}% Include figure files
\usepackage{amssymb}
\usepackage[english]{babel}
 \usepackage{amsthm}
 \usepackage[utf8]{inputenc}
\usepackage{bm}
\usepackage{dcolumn}% Include figure files
\usepackage{latexsym} 

\newcommand{\mbfx}{{\bf x}}
\newcommand{\mbfw}{{\bf w}}
\newcommand{\mbfa}{{\bf a}}
\newcommand{\mbfb}{{\bf b}}
\newcommand{\mbfg}{{\bf g}}
\newcommand{\mbfh}{{\bf h}}

\begin{document}

%\draft
\title{Theory of neuromorphic computing  by waves:
\\machine learning by rogue waves, dispersive shocks, and solitons}
\author{Giulia Marcucci}
\author{Davide Pierangeli}
\author{Claudio Conti}
\affiliation{
  Institute for Complex Systems, National Research Council (ISC-CNR),
  Via dei Taurini 19, 00185 Rome, Italy}
\affiliation{
 Department of Physics, University Sapienza, Piazzale Aldo Moro 2, 00185, Rome, Italy}
\email{claudio.conti@uniroma1.it}
\date{\today}

\begin{abstract}
  We study artificial neural networks with nonlinear waves as a computing reservoir. We discuss universality and the conditions to learn a dataset in terms of output channels and nonlinearity. A feed-forward three-layer model, with an encoding input layer, a wave layer, and a decoding readout, behaves as a conventional neural network in approximating mathematical functions, real-world datasets, and universal Boolean gates. The rank of the transmission matrix has a fundamental role in assessing the learning abilities of the wave. For a given set of training points, a threshold nonlinearity for universal interpolation exists. When considering the nonlinear
  Schr\"odinger equation, the use of highly nonlinear regimes implies that solitons, rogue, and shock waves do have a leading role in training and computing. Our results may enable the realization of novel machine learning devices by using diverse physical systems, as nonlinear optics, hydrodynamics, polaritonics, and Bose-Einstein condensates. The application of these concepts to photonics opens the way to a large class of accelerators and new computational paradigms. In complex wave systems, as multimodal fibers, integrated optical circuits, random,
  topological devices, and metasurfaces, nonlinear waves can be employed to perform computation and solve complex combinatorial optimization.\end{abstract}

%\pacs{}

\maketitle
\begin{figure}
 \includegraphics[width=\columnwidth]{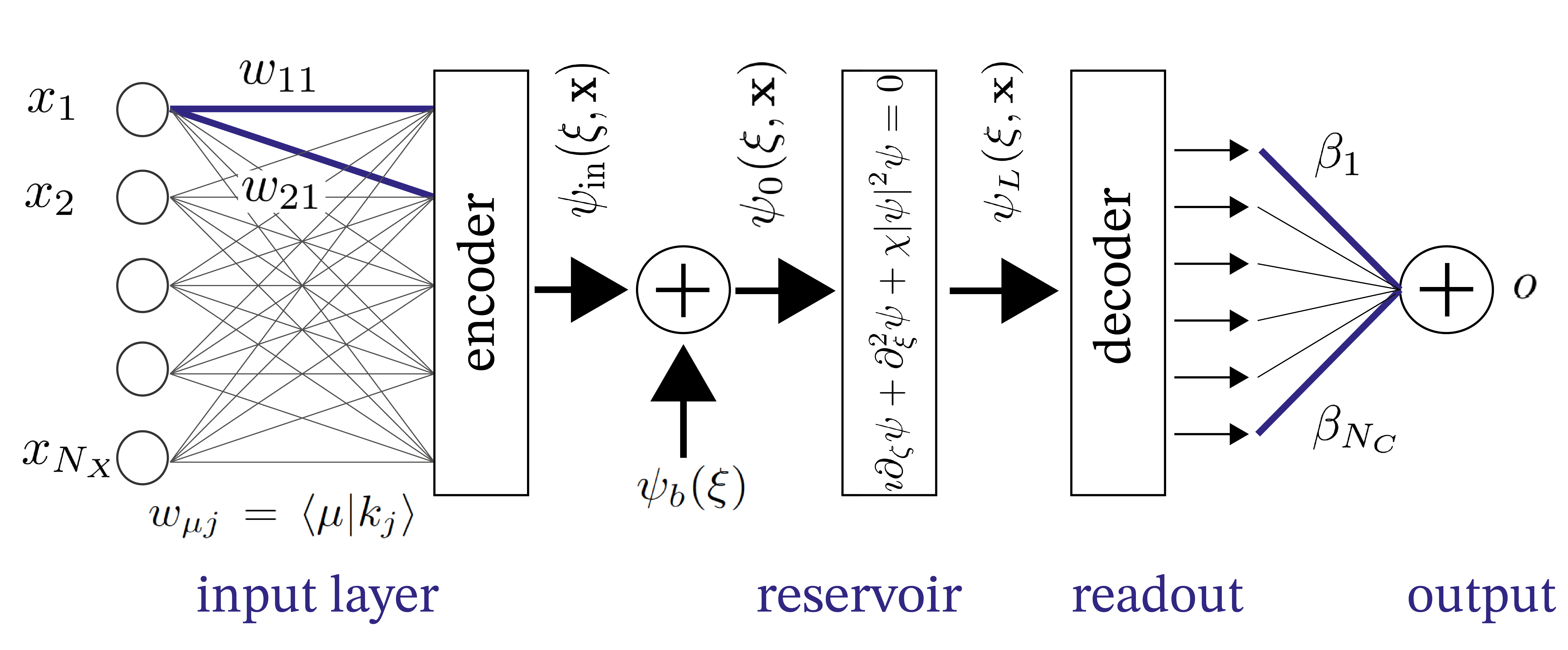}
  \caption{
    Single wave-layer feed-forward neural network.
    The input vector $\mbfx$ is encoded in input wave $\psi_0$, including
    a bias wave function $\psi_b$. The wave evolves according to a nonlinear partial differential equation.
    The readout layer decodes by sampling the modulus square $|\psi_L|^2$
    of the final wave in $N_C$ readout channels, linearly combined to give the output $o$.}
  \label{fig:scheme}
\end{figure}

Computational models as deep artificial neural networks have unprecedented success in learning large datasets for, e.g.,  image classification or speech synthesis~\cite{LeCun2015, Schmidhuber2015}. However, when the number of weights grows, model optimization becomes hard. Less demanding computational architectures exist, as neuromorphic and random neural networks, where the training involves only a subset of nodes~\cite{Pao1994,Jaeger2004,Huang2006,Verstraeten2007,Widrow2013,Wang2017,Scardapane2017,Gallicchio2018}. As models depart from conventional deep learning, many physical systems suit to large-scale computing~\cite{Ott2018}. Following earlier investigations~\cite{Psaltis1985, DenzBook}, various groups reported on computing machines with propagating waves,
like  Wi-Fi waves~\cite{Hougne2018}, polaritons~\cite{ballarini2019polaritonic, Opala2019}, and lasers
\cite{Duport2012,Vandoorne2014,Englund2017,Van2017,Bueno2018,Fratalocchi2018,Zhao2018, Engheta2019,Dong2020,Ludge2020}. The photonic accelerators speed up large-scale neural networks~\cite{Brunner2013, Shen2017, Lin2018, Pleros2019} or Ising machines~\cite{Wu2014,Marandi2016,Roques-Carmes2018,Kalinin2018,Pierangeli2019,Boehm2019,Takesue2019}. Despite these many investigations, fundamental questions remain open. {\it What kind of computation can waves do?} {\it Are linear or nonlinear waves universal computing machines?} {\it Which parameter signals the onset
  of the learning process?}
We need a general theory that links the physics of nonlinear waves
with the maths of machine learning and reservoir computing~\cite{Rudy2017}.
Also, no theoretical work addresses highly nonlinear processes, as nonlinear gases, shocks, and solitons (recently subject to intense research~\cite{KivsharBook,Barland02,Genty2018,Turitsyn2019a,Marcucci2019a,Bonnefoy2019,Naveau2019}), to perform computation at a large scale.
In this Letter,  we study computational machines with layers of nodes replaced by nonlinear waves. We discuss the conditions for learning, and demonstrate function interpolation, datasets, and Boolean operations.
\\{\it The wave-layer model ---} Figure~\ref{fig:scheme} shows our model, a ``single wave-layer feed-forward network'' (SWFN), in
analogy with the ``single layer feed-forward neural network'' (SLFN)
\cite{Schmidhuber2015,Huang2006,Jaeger2004}. An input vector ${\bf x}\in\mathbb{R}^{N_X}$ seeds the network, which has $N_C$ output {\it channels} $g_j$ with  $j=1,2,...,N_C$ forming the output vector ${\bf g}\in\mathbb{R}^{N_C}$.
The channels are linearly combined at the readout $o=\sum _{j=1}^{N_c} \beta_j g_j=\bm{\beta}\cdot{\bf g}$. $\bm{\beta}\in\mathbb{R}^{N_C}$ is the readout vector with weights $\beta_1,\beta_2,...,\beta_{N_C}$ determined by the training dataset. The link between ${\bf g}$ and ${\bf x}$ is given by a scalar complex-valued field $\psi_L(\xi,\mbfx)$ emerging from the propagation of an initial field $\psi_0(\xi,\mbfx)$, with $\xi$ the {\it transverse coordinate}.
In our simplest formulation, the readout functions are samples of
 $|\psi_L(\xi,\mbfx)|^2$ in a finite set of points  $\bar \xi_j$, with $j=1,2,...,N_T$, i.e., $g_j=|\psi_L(\bar\xi_j,\mbfx)|^2$ such that
  \begin{equation}
    o(\mbfx)=\sum_{j=1}^{N_C} \beta_j |\psi_L(\bar\xi_j,\mbfx)|=\sum_{j=1}^{N_C} \beta_j g_j(\mbfx)=\bm{\beta}\cdot\mbfg(\mbfx).
    \end{equation}
The goal is determining the weights $\bm{\beta}$ to approximate a target function $f(\mbfx)$ or learn a finite dataset. The final state $\psi_L(\xi,\mbfx)$ depends on the input vector $\mbfx$, because $\mbfx$ is encoded in the initial condition $\psi_0(\xi,\mbfx)$. Using Dirac brackets to simplify the notation, we have $\psi_0(\xi,\mbfx)=\langle \xi | 0,\mbfx\rangle,\mbfx$ at $\zeta=0$, and the final state is $\psi_L(\xi,\mbfx)=\langle \xi | L,\mbfx\rangle$ at $\zeta=L$. $\zeta$ is the {\it evolution coordinate}.  The evolution from $\psi_0(\xi,\mbfx)$ to $\psi_L(\xi,\mbfx)$ follows a nonlinear partial differential equation.
    There is no specific receipt for the equation, one can also consider nonlinear equation with external random or deterministic potentials.
    We use the nonlinear Schr\"odinger equation
\begin{equation}
  \label{eq:NLS}
  \imath \partial_\zeta\psi+\partial_\xi^2\psi+\chi |\psi|^2 \psi=0\text{,}
\end{equation}
with $\psi(\xi,\zeta=0)=\psi_0(\xi,\mbfx)$, and $\psi(\xi,\zeta=L)=\psi_L(\xi,\mbfx)$. In~(\ref{eq:NLS}), $\chi$ measures the strength of the nonlinearity, and $\chi=0$ corresponds to linear propagation.
We write the input state as $|0,\mbfx\rangle=|\text{in},\mbfx\rangle+|b\rangle$, where $|b\rangle$ is the {\it bias state} described below (see Fig.~\ref{fig:scheme}). The encoder maps $\mbfx$ in $|\text{in},\mbfx\rangle$ in the Fourier (or momentum) space (this follows a typical approach in optics with a spatial light modulator).
Given a set of plane waves $|k_j \rangle$ with $j=1,2,...,N_X$, we have
  \begin{equation}
  |\text{in},{\bf x}\rangle=\sum_{j=1}^{N_X}x_j |k_j\rangle,
  \end{equation}
  with $\psi_{\text{in}}(\xi,\mbfx)=\langle \xi |\text{in},\mbfx\rangle$.
  For the initial state we have
  \begin{equation}
    \psi_0(\xi,\mbfx)=\langle \xi |0,\mbfx\rangle= \sum_{j=1}^{N_X}x_j \langle\xi|k_j\rangle+\langle\xi|b\rangle.
    \end{equation}
The states can be represented by a generalized basis $|\mu\rangle$:
  \begin{equation}
    |0,\mbfx\rangle=\sum_{\mu}a_{\mu}|\mu\rangle=
\sum_\mu\left(\sum_{j=1}^{N_X}w_{\mu j}x_j+b_{\mu}\right) |\mu\rangle. \end{equation}
The input state is the vector $\mbfa$, with components $a_\mu$
\begin{equation}
  \label{eq:inic}
  a_{\mu}=\sum_{j=1}^{N_X}w_{\mu j}x_j+b_{\mu} ,
\end{equation}
$w_{\mu j}=\langle \mu|k_j\rangle$ are the input weights, and $b_{\mu}=\langle \mu| b\rangle$. In the finite-bandwidth case, the wave function can be uniquely determined by a numerable set of samples, according to the Nyquist-Shannon theorem, being $|\mu\rangle=|\xi_{\mu}\rangle$, with $\mu$ a discrete index, and $\xi_\mu$ sampling points.
\\{\it Training ---}
The model maps the SWFN training to conventional reservoir computing protocols. Given a finite number $N_T$ of training points $\mbfx^{(t)}$, with $t=1,2,...,N_T$, and corresponding targets $T^{(t)}$, the wave ``learns''
the training set when $o^{(t)}=o[\mbfx^{(t)}]=T^{(t)}$ for any $t$.

\begin{figure}
\includegraphics[width=\columnwidth]{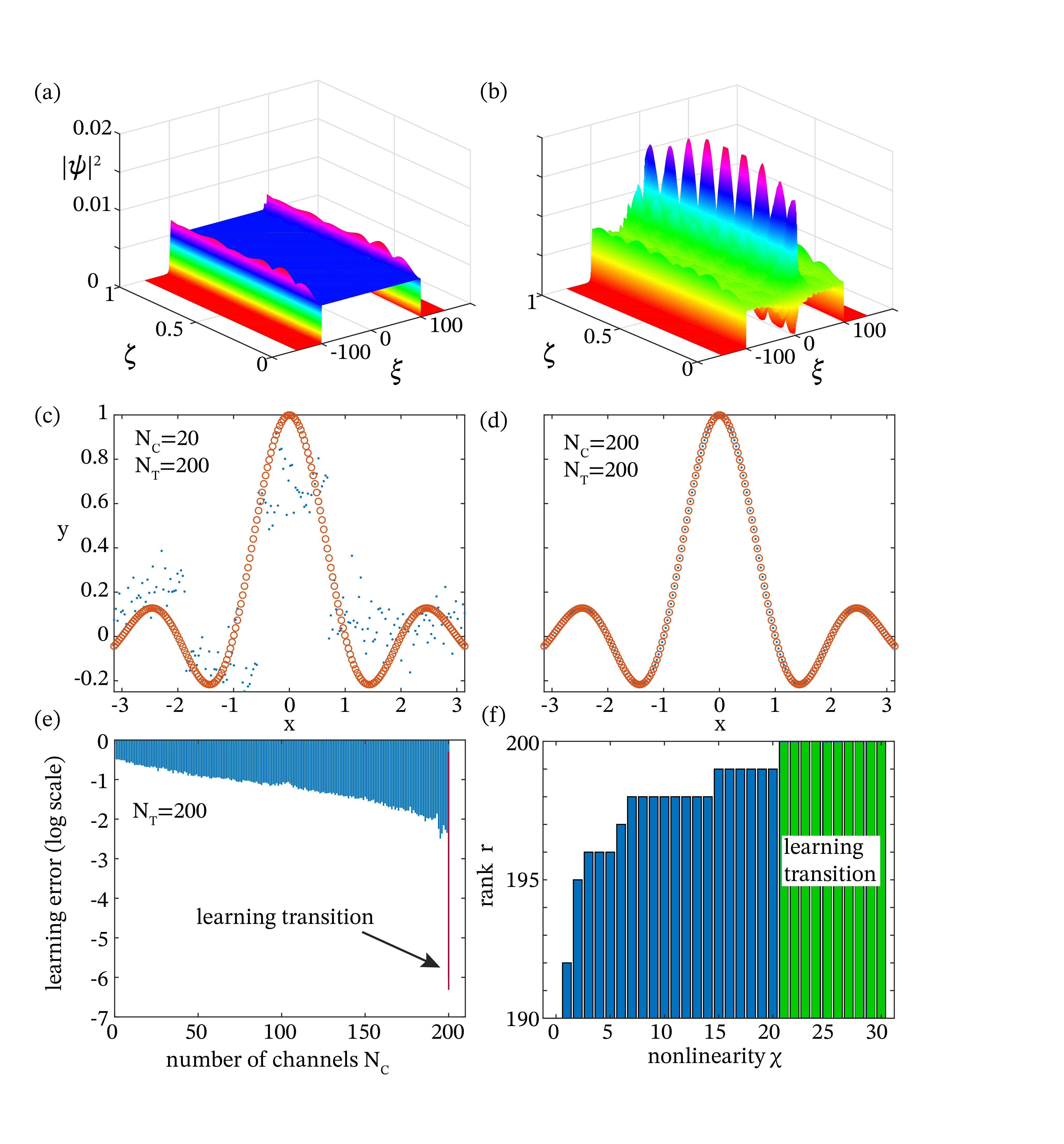}
  \caption{
    Learning the function $y=\sin(x)/x$:
    (a) Bias evolution; (b) wave evolution for a representative point $x=-\pi$;
    (c) training points (red circles) and
    SWFN output (dots) after training with $N_T=200$ and $N_C=20$; (d) as in (c) for
    $N_T=200$ and $N_C=200$;
    (e) training error (log-scale) when increasing $N_C$ at fixed nonlinearity $\chi=25$,
    the learning threshold at $N_C=N_T$ is evident; (f) rank $r$
    of the transmission matrix $\bm{H}$ varying the nonlinearity $\chi$
    for $N_C=N_T=200$, the threshold $\chi$ for learning corresponds to $r=N_T$.}
  \label{fig:sinc}
\end{figure}

\begin{figure}
  \includegraphics[width=\columnwidth]{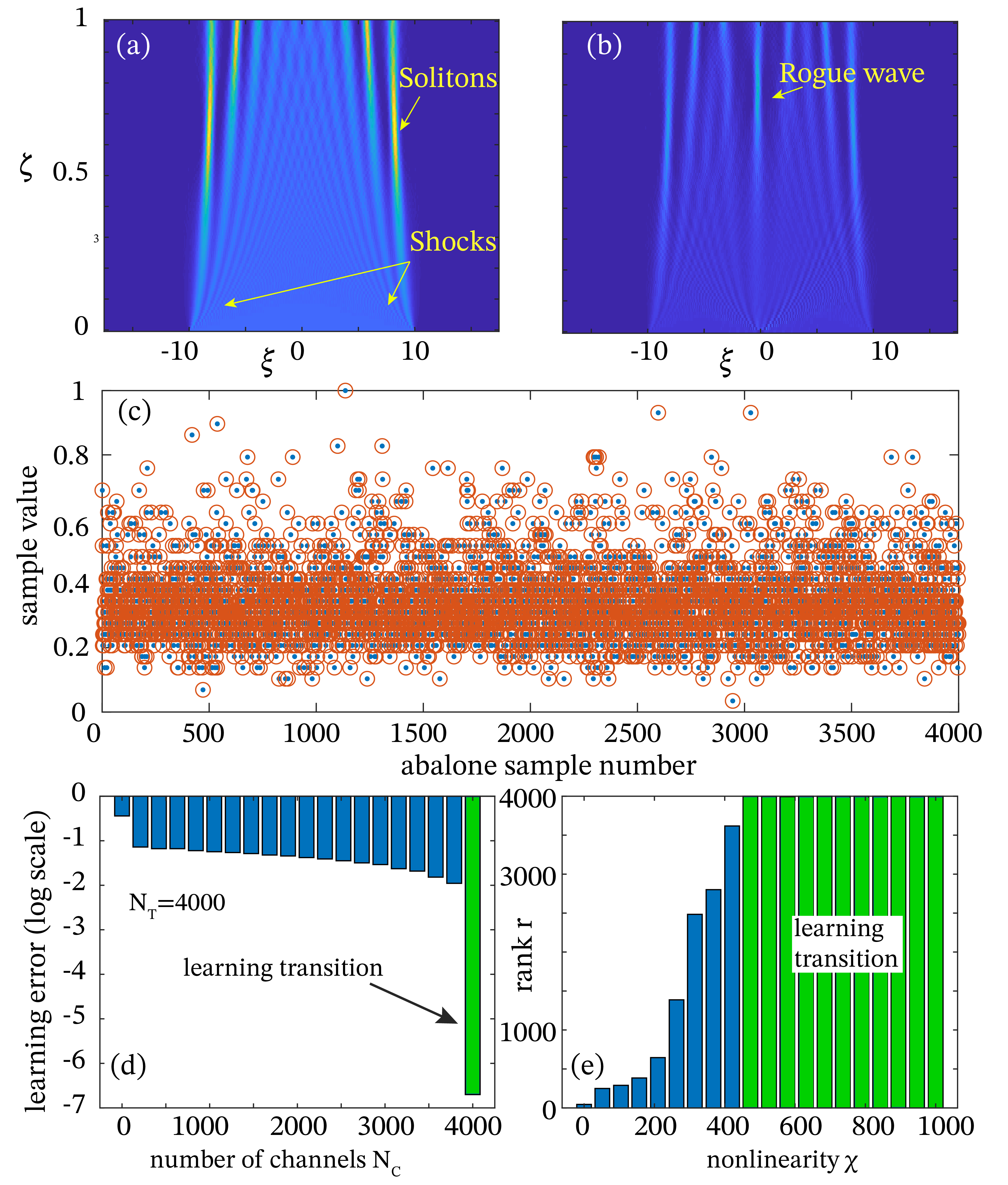}
  \caption{
    Learning of the abalone dataset: (a) Bias function encompassing two
    counter-propagating shock waves and soliton generation;
    (b) propagation of one representative point of the dataset with the generation of a rogue wave; (c) comparison between training data (circle)
    and SWFN outputs (dots), with $N_T=N_C=4000$, and $\chi=500$, showing that the SWFN has learned the data;
    (d) training error for $N_T=4000$ and $\chi=500$ Vs $N_C$; (e) rank $r$ Vs nonlinearity $\chi$ for $N_T=N_C=4000$ showing
  the learning transition.}
  \label{fig:abalone}
\end{figure}

\begin{figure}
  \includegraphics[width=\columnwidth]{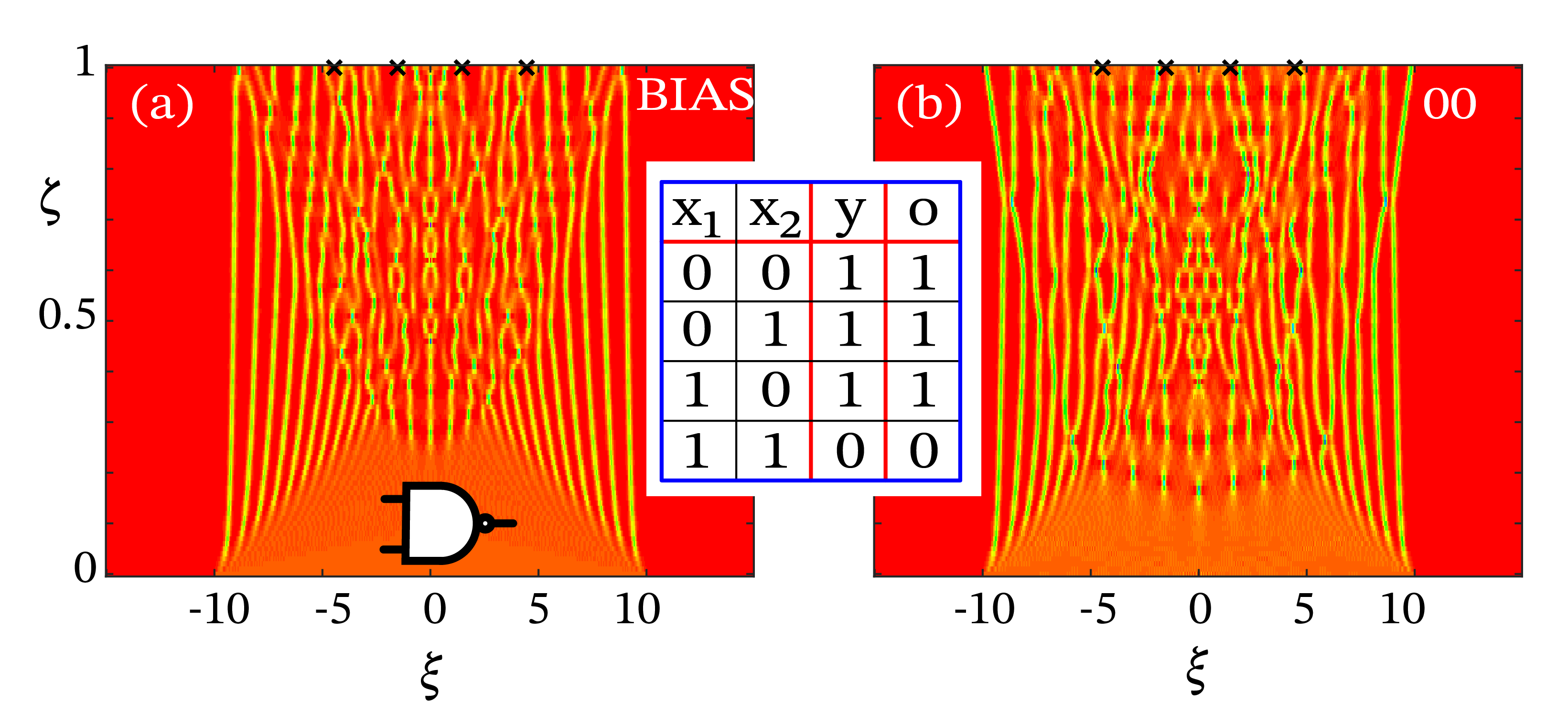}
  \caption{Training a universal logic gates by using shock waves and soliton gases
    ($\chi=1000$): (a) Bias function with two counter-propagating shocks in the
    initial dynamics evolving into solitons and rogue waves;
    (b) propagation of the input $\bm{0}\bm{0}$ in the NAND gate.
    The crosses at $\zeta=1$ correspond to the readout sampling points ($N_C=N_T=4$).
    The inset shows the NAND gate truth table for training, and the SWFN output $o$.
  \label{fig:nand}}\end{figure}

  To establish the conditions for learning, we evaluate the $N_T\times N_X$ {\it transmission matrix} $H_{tj}=g_j[\mbfx^{(t)}]=|\psi_L(\bar \xi_j,\mbfw\cdot\mbfx_t+\mbfb)|^2$ by evolving the wave on all the $N_T$ inputs $\psi_0(\xi,\mbfx^{(t)})$, and solve the linear system
\begin{equation}
  o(\mbfx_t)=\sum_{j=1}^{N_C} H_{tj}\beta_j=f(\mbfx_t)=T^{(t)},
  \label{eq:Ho}
  \end{equation}
  or $\bm{H}\cdot\bm{\beta}=\bm{T}$, with $\bm{T}\in\mathbb{R}^{N_T}$ with components $T^{(t)}$.  The machine can learn the entire dataset with zero error only if $N_C=N_T$ and $\bm{H}$ has rank $N_C$. In this case, $\bm{\beta}$ exists such that $||\bm{H}\cdot \bm{\beta}-\bm{T}||=0$. In the general case $N_C\leq N_T$, the smallest norm least squares solution is found by the Moore-Penrose pseudo-inverse of $\bm{H}$~\cite{Huang2006}.   
  \\
  However, even when $N_C=N_T$, the matrix $\bm{H}$ is not always invertible, and hence the model cannot learn the dataset with arbitrary precision. We find that there are two conditions for learning: (i) the function $|\psi_L(\bar \xi_j,\mbfw\cdot\mbfx_t+\mbfb)|^2$ must be non polynomial in $\mbfb$, and (ii) in $\xi$.
  The need for condition (i) arises from Eq.~(\ref{eq:Ho}), since $f(\mbfx)$ is an arbitrary function, and can only be represented by non polynomial functions~\cite{Hornik1990,Leshno1993}.  
  The need for condition (ii) can be demonstrated following Refs.~\cite{Tamura1997,Huang2006}: the vector $\mbfh(\xi)$ with components $h_t(\xi)=|\psi_L(\bar \xi,\mbfw\cdot\mbfx_t+\mbfb)|^2$
  spans $\mathbb{R}^{N_T}$ if $|\psi_L(\xi,\mbfw\cdot\mbfx_t+\mbfb)|^2$ and
  all its derivatives w.r.t. $\xi$ are non-vanishing.
  Correspondingly, $N_T$ points $\bar\xi_j$ exist,
  such that $\mbfh(\bar\xi_j)$ is a basis, and $\bm{H}$ has rank $r=N_T$.
Notably, condition (ii) is always satisfied in the cases of interest here, as $|\psi_L(\xi,\mbfw\cdot\mbfx_t+\mbfb)|^2$ is a probability density w.r.t. $\xi$, within a normalization factor.  
\\\noindent Conditions (i) and (ii) have two significant consequences:
(I) If the wave evolution is linear, the wave {\it is not} a universal approximator, indeed, $o(\mbfx)$ is only a quadratic function of $\mbfb$. (II) {\it Not all} nonlinear evolutions act as universal approximators. For example, a wave that undergoes only self-phase modulation, i.e.,
$\psi_L(\xi)=\psi_0(\xi)\exp[\imath |\psi_0(\xi)|^2 L]$, will not result in a non polynomial output function.
The same holds for a weakly nonlinear (perturbative) evolution. To have universal approximators, one needs that the final wave is a strongly nonlinear function of the input, which means that all the derivatives w.r.t. the bias wave parameters (amplitude, width, etc.) must be non-vanishing. In other words, starting from a linear propagation [$\chi=0$ in Eq.~(\ref{eq:NLS})], when augmenting the nonlinearity $|\chi|$, one observes an increasing ability to learn, corresponding to a growing rank $r$. For a given training set, there is a threshold nonlinearity for the training error to be zero.
\\\noindent To test these arguments, we use numerical simulations. In the examples below, we show:
(i) The SWFN can approximate arbitrary functions and learn datasets as conventional reservoir computing only above a critical nonlinearity. (ii) Linear propagation does not act as a universal approximator. (iii) The SWFN can implement universal Boolean logic gates, as the NAND. (iv) The SWFN can perform with binary and real valued inputs.
\\{\it Example 1, $\sin(x)/x$ with binary encoding ---}
We first show learning input/output functions.
We consider $y=\sin(x)/x$, and adopt a binary encoding of the real variable $x$ in the range $[-\pi,\pi]$:
we quantize $x$ by $N_X=12$ bits, such that the input is a string of $\bm{0}$ and $\bm{1}$.
We use $N_X$ plane waves $|k_j\rangle$ with unitary amplitude and phase $0$ for $\bm{1}$ and $\pi$ for $\bm{0}$.
In the simulation of Fig.~\ref{eq:NLS}, $\xi$ is discretized with
$512$ points in the domain $[-150,150]$, and $L=1$.
The  $N_C$ channels are linearly distributed in the range $[-100,100]$.
The bias is a rectangular wave with amplitude $a_b=1$ and half width $w_b=100$ (Figure~\ref{fig:sinc}a).
Figure~\ref{fig:sinc}b shows the evolution of a representative $\mbfx$. We study the learning when increasing $N_C$, at $N_T=200$ and $\chi=25$.
Figure~\ref{fig:sinc}c illustrates the training data compared with the SWFN output for $N_C=20$ and $\chi=25$; Fig.~\ref{fig:sinc}d the case for $N_C=N_T=200$. In the latter case, all the training points are learned with zero error within numerical precision.
Exact learning occurs at $N_C=N_T$, when the error abruptly drops of several order of magnitudes, this is evident in Fig.~\ref{fig:sinc}e, showing the training error versus $N_C$.
%%%%%%%%%%%%%%%%%%%%%%%%%%%%%%%%%%%%%%%%%%%%%%%%%%%%%%%%%
%%%%%%%%%%%%%%%%%%%%%%%%%%%%%%%%%%%%%%%%%%%%%%%%%%%%%%%%%
The role of nonlinearity is studied in Fig.~\ref{fig:sinc}f.
Learning requires that the rank $r$ of $\bm{H}$ is equal to $N_C$.
Fig.~\ref{fig:sinc}f shows $r$ versus $\chi$ for $N_C=N_T=200$.
Only above a threshold value for $\chi$, the learning condition $r=N_T$ is achieved.
%%%%%%%%%%%%%%%%%%%%%%%%%%%%%%%%% EXAMPLE 2
\\{\it Example 2, abalone dataset with amplitude encoding ---}
We test learning conventional datasets for neural networks.
We consider the ``abalone dataset'' \footnote{https://archive.ics.uci.edu/ml/datasets/Abalone}. Each training input has $N_X=8$, and we use $N_T=4000$.
Figure~\ref{fig:abalone}a,b show the bias wave function and a representative $\mbfx$ propagation.
Figure~\ref{fig:abalone}c shows values from the training set (red circles) and the SWFN output (dots) for $N_C=4000$ and $\chi=500$. The learning transition is shown when varying $N_C$ for $\chi=500$ in Fig.~\ref{fig:abalone}d; and varying $\chi$ in Figure~\ref{fig:abalone}d for $N_C=N_T=4000$. Learning a large datasets needs a nonlinear propagation.
The various inputs $\mbfx$ generate different
ensembles of nonlinear waves, as shocks, solitons and rogue waves,
resembling recent experimental results~\cite{Marcucci2019a}.
%%%%%%%%%%%%%%%%%%%%%%%%%%%%%%%%%%%%%%
%%%%%%%%%%%%%%%%%%%%%%%%%%%%%%%%%%%%%%
\\{\it Example 3, Boolean logic gate ---}
The SWFN can also realize universal classical Boolean logic gates.
We consider for example the NAND gate, which has two binary inputs and
one binary output with the corresponding truth table in Figure~\ref{fig:nand}.
We encode the two inputs in the phases of two plane waves in the $k-$space.
Being $N_X=2$, $x_j=1$ (phase $0$) correspond to the binary symbol $\bm{0}$,
and $x_j=-1$ (phase $\pi$)for the binary symbol $\bm{1}$.
Figure~\ref{fig:nand} shows the resulting trained gate.
The SWFN output is given in the truth table and obtained
at the machine precision within an error of $10^{-15}$.
The rectangular bias evolution is shown in Fig.~\ref{fig:nand}b.
The sampling points $\bar\xi_{1,2,3,4}$ at the readout are also indicated.
Figures~\ref{fig:nand}a,b show two wave propagations: Two counter-propagating shock waves form in the bias evolution (Fig.~\ref{fig:nand}a). The input vector ${\bm 0}{\bm 0}$ superimposed to the bias highly nonlinear bias produces the needed output (Fig.~\ref{fig:nand}b). Many solitons and rogue waves are visible.
The neuromorphic nonlinear wave device learns the truth table and performs
the logical operation. Similar results are obtained for other Boolean logic gates.
\\{\it Conclusions ---}
We studied theoretically artificial neural networks with a nonlinear wave as a reservoir layer. We found that they are universal approximators, and developed a new computing model driven by nonlinear partial differential equations. Following general theorems in neural networks,  we discovered that a threshold nonlinearity exists for the learning process. When the dataset is large, the threshold implies the generation of highly nonlinear processes as dispersive shocks, rogue waves, and soliton gases. The rank of the transmission matrix is the relevant parameter to assess the learning transition when varying the channels and the degree of nonlinearity. Linear or weakly nonlinear evolution may only learn small datasets.

Our results may stimulate further theoretical work to determine the best nonlinear phenomena for learning, and even their ability to generalize. The roles of external potentials, randomness, and noise in quantum and turbulent regimes are unexplored. Our framework holds in optics, polaritonics, hydrodynamics, Bose-Einstein condensates, and all the fields encompassing nonlinearity, also with models different from the nonlinear Schroedinger equation. Our analysis is the starting point for many other developments, as cascading wave-layers and conventional nodes. If building heterogeneous deep computational systems with standard neural networks and waves provides computing advantages is an open question. Photonics speeds up electronic systems for machine learning; our theoretical results foster the development of new computing hardware.

The present research was supported by PRIN 2015 NEMO project (grant number 2015KEZNYM), H2020 QuantERA QUOMPLEX (grant number 731473), H2020 PhoQus (grant number 820392), PRIN 2017 PELM (grant number 20177PSCKT), and  Sapienza Ateneo.


\end{document}